\documentclass[preprint,eqsecnum,secnumarabic,aps,showpacs]{revtex4}

\usepackage{array}
\usepackage{delarray}
\usepackage{tabularx}
\usepackage{hhline}
\usepackage{dcolumn}
\usepackage{graphicx}

\newcommand{\PreserveBackslash}[1]{\let\temp=\\#1\let\\=\temp}
\let\PBS=\PreserveBackslash

\begin{document}

\date{\today}

\topmargin +0.5cm

\title {Exchange Bias Theory: a Review}

\author{Miguel Kiwi} 

\affiliation
{Facultad de F\'{\i}sica, Pontificia Universidad Cat\' olica de
Chile,\\ 
Casilla 306, Santiago, Chile  6904411}

\begin{abstract} 

Research on the exchange bias (EB) phenomenon has witnessed a flurry
of activity during recent years, which stems from its use in magnetic
sensors and as stabilizers in magnetic reading heads. EB was
discovered in 1956 but it attracted only limited attention until
these applications, closely related to giant magnetoresistance, were
developed during the last decade. In this review I first give a short
introduction, listing the most salient experimental results and what
is required from an EB theory. Next, I indicate some of the obstacles
in the road towards a satisfactory understanding of the phenomenon.
The main body of the text reviews and critically discusses the
activity that has flourished, mainly during the last five years, in
the theoretical front. Finally, an evaluation of the progress made,
and a critical assessment as to where we stand nowadays along the
road to a satisfactory theory, is presented.  
\end{abstract}

\vskip 2.cm

\maketitle
\newpage 

\section{Introduction}
\label{introduction}

A complete theoretical understanding of the exchange bias (EB)
phenomenon has posed a formidable challenge to condensed matter
theorists for over four decades. The challenge emanates from several
sources: the intrinsic interest of EB, the many supplementary
physical phenomena that are involved and the important technological
applications that have been developed recently.  EB was discovered
almost half a century ago, by Meiklejohn and Bean~\cite{meik56}, and
its characteristic signature is the shift of the center of magnetic
hysteresis loop from its normal position at H$=0$ to H$_E\neq 0$. It
occurs in a large variety of systems~\cite{nogu99_rev} which are
composed by an antiferromagnet (AF) that is in atomic contact with a
ferromagnet (F) if the sample is grown, or after the system is
cooled, below the respective N\'eel and Curie temperatures $T_N$ and
$T_C$, in an external cooling field H$_{cf}$. Examples of the type of
systems where EB has been observed are clusters or small particles, F
films deposited on single crystal or polycrystalline AF's and F/AF
thin films bilayers, and spin glasses.  A comprehensive review, which
emphasizes experimental results and provides an up-to-date list of
relevant publications, was recently published by Nogu\'es and
Schuller~\cite{nogu99_rev}.  While I will not ignore experimental
observations, I refer the interested reader to Ref.~\cite{nogu99_rev}
for extensive and detailed information. Other reviews have also been
published, which discuss both theory and experiment, by Berkowitz
and Takano~\cite{berk99} and very recently a paper by
Stamps~\cite{stam00}, which includes novel results.

Defining as positive the direction of the cooling field H$_{cf}$, in
general the EB shift is towards negative fields, {\it i.e.} H$_E <
0$; however, recently Nogu\'es {\it et al.}~\cite{nogu96} found that
samples exposed to large cooling fields (H$_{cf}\sim 1$~Tesla) can
exhibit positive EB, {\it i.e.} H$_E > 0$.

Several supplementary remarkable features are associated with EB, in
addition to the symmetry breaking related to the appearance of the
{\it unidirectional} anisotropy that brings about H$_E\neq 0$.  Among
them is the existence of a blocking temperature $T_B$ above which EB
vanishes. While usually $T_B\approx T_N$ ({\it e.g.} F slabs grown on
the (111) face of NiO~\cite{shen96}) $T_B$ can be considerably lower
than the N\'eel temperature ({\it e.g.} AF's obtained through
oxidation of permalloy~\cite{char71,fulc72}).  Another remarkable
feature of EB is the training effect, {\it i.e.} the dependence of
H$_E$ on the number of measurements $n$, with the value of H$_E$
decreasing as $n$ increases~\cite{char71,fulc72,tsan82}, which
constitutes a hint that the interface actually is in metastable
equilibrium.  More recently, an important additional feature was
discovered: the memory effect, which consists in the fact that the
system keeps a memory of the temperature at which it was
field-cooled~\cite{ambr97,chou98,wu98,goke99,goke99b,li00}. The latter is
closely related to the freezing of the AF magnetic structure reported
by Ball {\it et al.}~\cite{ball96}. Still another characteristic
associated with many EB systems, observed when $T<T_B$ and which
appears to have a magnetic origin~\cite{leig00,hou00}, is a large
increase of the coercivity. 

As emphasized in the comprehensive review by Nogu\'es and
Schuller~\cite{nogu99_rev}, the EB phenomenon has recently received
renewed attention due to its important technological applications.
However, and in spite of this renovated interest, a full
understanding of EB has not yet emerged.  In this review I briefly
mention some of the major experimental results that have been
obtained over the years, but my attention is focused on the present
status of the theoretical understanding of the phenomenon.  The
dynamic pace with which the field has developed, especially during
the last five years, precludes me from mentioning every relevant
contribution that has been accomplished recently; I can only ask the
unjustly ignored authors for their indulgence.

Historically, the phenomenon was first observed in fine Co particles
covered by CoO~\cite{meik56}; this case forms part of the family of
systems of small particles coated by their native oxide (like
Ni/NiO~\cite{yao96} and Fe/Fe$_3$O$_4$~\cite{papa90}) or by their
nitride (like Co/CoN~\cite{lin95} and Fe/Fe$_2$N~\cite{hsu94}),
although it has also been observed in hybrid systems like
Co/NiO~\cite{sort99}.  However, I will concentrate on the most widely
studied group, namely EB materials in the form of thin film bilayers.
There are several reasons for this preference: i)~experimentally these
systems allow the best possible control and characterization of the
interface~\cite{nogu99_rev,nogu96b,jung94}; ii)~most of the actual
devices based on EB properties are fabricated in this
form~\cite{dien91a,dien91b,dien95}; and, iii)~these systems are the
most convenient to carry out the investigation of specific properties,
such as the role of interface structure~\cite{nogu99,goke01},
interface magnetic coupling direction~\cite{mora98,ijir98,nogu00},
cooling field intensity dependence~\cite{nogu96,leig99}, large
EB~\cite{nogu96b}, coercivity enhancement~\cite{leig00,hou00}, and the
deviations from inversion symmetry of the hysteresis
loop~\cite{fitz00}, among others. All the preceding items are relevant
when trying to develop a proper theoretical description and
understanding of EB.

An additional relevant characteristic of the interface is best
defined at this stage: the important distinction between compensated
and uncompensated AF interface layers. In the former the net total
magnetic moment of the AF interface layer vanishes, since the vector
addition of the spins that belong to each of the two AF magnetic
sublattices, and point in opposite directions, cancels out.
Conversely, in an uncompensated AF magnetic face all spins point in
the same direction and the layer has a net magnetic moment.  These
definitions become quite relevant when I examine the early theories
put forward to describe the interface properties of a F/AF system.

It is now appropriate to formulate precisely what I mean by a proper
theoretical description and understanding of EB. The list of
requirements that I define below certainly is neither unique nor all
one could wish for, but constitutes the minimal knowledge a theorist
would like to have for oneself and to supply an experimentalist with.
However, this minimum still is far in the horizon of accomplished
achievements.  My restricted list reads as follows: 1)~first and
foremost, to set down a mechanism, {\it free of ad hoc assumptions on
the interface structure}, that yields the genuine reason for unidirectional
anisotropy; 2)~to derive values for the magnitude of the exchange
anisotropy field H$_E$ and the coercivity H$_c$, as functions of the
temperature $T$ and the cooling field H$_{cf}$; 3)~to understand why
compensated interfaces yield values of H$_E$ larger, or at least as
large, as uncompensated interfaces; 4)~to understand the highly
nontrivial relation between interface roughness and EB; 5)~to explain
the memory effect and how it is related to the blocking temperature;
6)~to explain the training effect; and, 7)~to understand the origin
of the inversion asymmetry that is often observed in the shape of the
shifted hysteresis loop.  

This review is organized as follows: after this Introduction, in
Sec.~\ref{interf_str}, I describe the main obstacle that blocks the
way, that is: the knowledge of the interface atomic and magnetic
structure. Next, in Sec.~\ref{theories}, I describe and classify a
set of different theoretical approaches that have been put forward.
Finally, in Sec.~\ref{final}, an evaluation of the present state of
affairs is given and conclusions are drawn.

\section{Interface structure: a hard nut to crack}
\label{interf_str}

The most relevant unknown element in the development of a
satisfactory understanding, and thus of a comprehensive theory of EB,
are the unknown features of the interface structure. On the one hand
the systems which exhibit EB are many and varied~\cite{nogu99_rev}:
thin films, single crystal AF with metallic coating, polycrystalline
and amorphous ferromagnets in contact with ordered and disordered AF
oxides and salts. On the other hand the interface, even in the most
ordered case of two single crystals in close contact, can have large
lattice parameter mismatches, strains and defects.  In addition, the
magnetic structure in the vicinity of the interface is not
necessarily identical to the bulk magnetic ordering~\cite{lede97}.

However, most generally the exact atomic arrangement in the vicinity
of the interface is either unknown or, at best, only quite uncertain.
Both crystallographic and magnetic relaxation and reconstruction might
develop at both sides of the interface and, to complicate matters
even further, these features are very rarely amenable to precise
experimental probing~\cite{char00}. 

The above obstacles are compounded by the complexity of the magnetic
structure, with many equivalent easy axes directions often present.
All in all this implies significant difficulties when trying to
formulate a sound theory. However, if one persists in trying to
achieve progress in developing a healthy theory of the EB phenomenon,
it is necessary to assume or postulate a definite single crystal 
and magnetic interface structure, which quite surely will not be
completely accurate nor will it incorporate all the subtle
intricacies of even the simplest systems~\cite{char00}.

\begin{figure} 
\begin{center}
\includegraphics[width=16.5cm,height=21.cm]{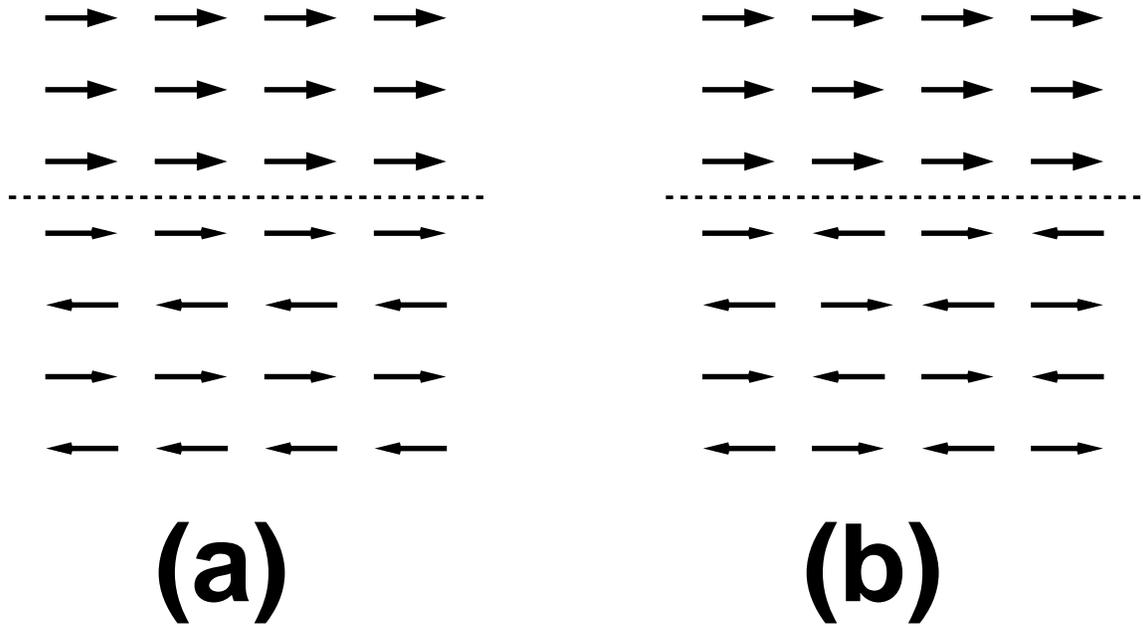}
\caption{Magnetically collinear AF uncompensated interface
  configuration. (a)~Ferromagnetic coupling across the interface:
  $J_{F/AF}>0$; (b)~AF coupling: $J_{F/AF}<0$.}
\label{fig:magn_colin} 
\end{center}
\end{figure}

\begin{figure} 
\begin{center}
\includegraphics[height=22cm]{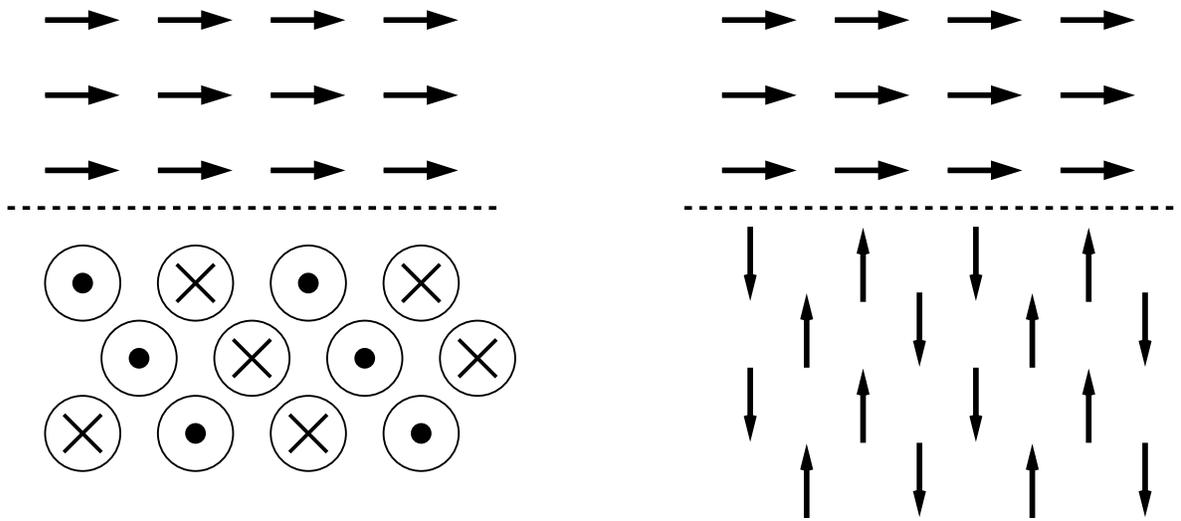}
\caption{Two examples of possible non-collinear
interface configurations.} 
\label{fig:magn_non-colin}
\end{center}
\end{figure}

Thus, it comes as no surprise that practically all the theories which
have been put
forward~\cite{neel67,neel88,mauri87,malo87,malo88a,malo88b,koon97,schu98,schu99,stil99,kmpr1,kmpr2,kmpr3,taka97,zhan99,dimi98}
at some point make a crucial assumption about the interface
crystallographic and magnetic structure.  For the time being I will
put aside the issue of the spatial atomic rearrangements
(reconstruction) in the vicinity of the interface and concentrate my
interest on the magnetic configuration of that region. Apart from the
trivial collinear interfacial magnetic structure shown in
Fig.~\ref{fig:magn_colin} there are many alternative structures, two
of which are illustrated in Fig.~\ref{fig:magn_non-colin}. They will
prove relevant to analyze the latest EB theoretical models.

\section{Theoretical models}
\label{theories}

In classifying and describing the attempts to develop a proper EB
theory, after a brief outline of earlier models, I will focus
attention on relatively recent work (publications after 1995).
Earlier models have in common the assumption of ground state
collinear magnetic structures on the F and AF side of the interface.
However, the actual interface structure is much richer and quite
complex, as was briefly described and discussed in
Sec.~\ref{interf_str}. At this point, and to help the reader with a
schematic outline of the present status of EB theories, I provide a
sketchy route map of what will be explored below. I give it the form
of a table that focuses on the main characteristics and results of
the various models that have been put forward.


\begin{tabular}
{|>{\PBS\raggedright\hspace{0pt}}p{3.0truecm}%
 |>{\PBS\raggedright\hspace{0pt}}p{3.2truecm}%
 |>{\PBS\raggedright\hspace{0pt}}p{4.5truecm}%
 |>{\PBS\raggedright\hspace{0pt}}p{4.6truecm}|
}

\toprule
{\bf Theory }&
{\bf Main features}&
{\bf Interface magnetic structure}&
{\bf Main result}\\  

\colrule\colrule

{\it Early work} ~\cite{meik62}
&Coherent F\&AF magnetization rotation 
&Uncompensated AF  interface layer; 
$\vec m_F \parallel \vec m_{AF}\ $ ($\vec m$: bulk magnetization)
& H$_E$ much larger than observed experimentally\\ 
\colrule
{\it N{\'e}el's model} ~\cite{neel67,neel88}
&Continuum approximation
&Uncompensated AF  interface layer;
$\vec m_F \parallel \vec m_{AF}$ 
& Domain wall in the AF; requires
large width of the F slab \\ 
\colrule
{\it Early random interface models} \cite{malo87,malo88a,malo88b}
&Random defects create random fields
&Uncompensated AF  interface layer; $\vec m_F \parallel \vec m_{AF}$
&Reasonable H$_E$ values which depend on defect concentration\\ 
\colrule
{\it AF domain wall models} ~\cite{mauri87}
&F interface coupling; thin F film
&Uncompensated AF  interface layer; $\vec m_F \parallel \vec
m_{AF}$&Reasonable H$_E$ values\\
\colrule
{\it Orthogonal F\&AF magnetization model}  \cite{koon97}
&Canting of the AF interface spins
&Compensated AF  interface layer; $\vec m_F \perp \vec m_{AF}\ $ 
&Realistic interface magnetic structure \\
\colrule
{\it Generalized random interface models} 
\cite{schu98,schu99,zhan99,dimi98}
&Rough interface; dipolar interaction is incorporated 
&AF interface compensated on average; $\vec m_F \perp \vec
m_{AF}$ and $\vec m_F \parallel \vec m_{AF}$ investigated
&Reasonable H$_E$ values; finite coercivity; dependent
on interface defect concentration\\ 
\colrule
{\it Frozen interface model} \cite{kmpr1,kmpr2,kmpr3}
&Spin glass like AF canted interface layer
&Compensated AF interface; $\vec m_F \perp \vec m_{AF}$
&Reasonable H$_E$ values; one adjustable parameter\\  
\colrule
{\it Local pinning field variation} \cite{stil99} 
&Full domain magnetization as basic element 
&Fluctuating easy axis directions of interface domains
&Reasonable values of H$_E$; finite coercivity; several adjustable
parameters\\   
\botrule
\end{tabular}

\newpage Inspection of the ``route map'' right away points out an
important feature: all the reviewed theories are based on simple
models, mainly the Ising and Heisenberg Hamiltonians. However, how
well these models describe actual interfaces, in particular the
metallic ones, is a debatable issue which was recently tested, using
first principles spin-density total energy methods, by Kurz~{\it et
al.}~\cite{kurz01}. 

While most of the theories that fit into one of the categories defined
in the ``route map'' are incompatible with those classified in the
rest of the categories, the significant influence of N\'eel's
contribution~\cite{neel67,neel88} on all work subsequent to his is
quite apparent. In his analysis he implemented analytically the
continuum approximation with several additional assumptions, but many
of the later papers, even those which based their calculations on
discrete treatments took, with more or less care as to their
applicability, his results for granted. Another important and
interesting milestone is the work by Koon~\cite{koon97}, which pointed
out the importance of the orthogonal magnetic configurations of the F
and AF magnetizations; while the model by itself does not yield EB it
strongly influenced later
work~\cite{schu98,schu99,stil99,kmpr1,kmpr2,kmpr3,taka97,zhan99,dimi98}.

I now proceed to explore and analyze in detail the theories roughly
classified in the ``route map''. 

\subsection{Early work}
\label{early_work}

The first attempt to develop an intuitive model for EB seems to be
due to Meiklejohn~\cite{meik62}. He assumed coherent rotation of the
magnetization F and the AF, and wrote for $\varepsilon$, the energy
per unit interface area, the following expression:

\begin{eqnarray}
    \varepsilon = &-& {\text H} M_F t_F \cos (\theta - \beta) 
      + K_F t_F \sin ^2 (\beta)\nonumber \\
      &+& K_{AF} t_{AF} \sin^2 (\alpha) - J_{F/AF}\cos (\beta
      -\alpha)\; .
    \label{eq:meik}
\end{eqnarray}

\noindent Above H is the applied magnetic field, $M_F$ the F
saturation magnetization, $t_f$ ($t_{AF}$) the thickness of the F
(AF) slab, $K_F$ ($K_{AF}$) the bulk anisotropy of the F (AF) and
$J_{F/AF}$ the interfacial exchange constant. The angles are defined
as follows: $\alpha$ is the angle formed by $\vec M_{AF}$ and the AF
anisotropy axis, $\beta$ is the angle formed by $\vec M_{F}$ and the
F anisotropy axis and $\theta$ is the angle between $\vec {\text H}$
and the F anisotropy axis. Neglecting the F anisotropy, which in
general is considerably smaller than $K_{AF}$, and minimizing respect
to $\alpha$ and $\beta$ the hysteresis loop shift
Meiklejohn~\cite{meik62} obtained is 

\begin{equation}
  {\text H}_E = \frac {J_{F/AF}} {a^2 M_F t_F}\; ,
  \label{eq:overestimate} 
\end{equation}

\noindent where $a$ is the lattice parameter. The order of magnitude
of ${\text H}_E$ that results depends on the unknown parameter
$J_{F/AF}$, a feature common to all of the theoretical models
developed in the EB context. Assuming $J_F \geq J_{F/AF} \geq J_{AF}$
the resulting value for H$_E$ is orders of magnitude larger than
experimentally observed~\cite{meik62}.  This overestimate is a
feature shared by several of the earlier models~\cite{neel67,meik62}.

It is interesting to point out that, if one adopts these earlier
models as a guide for an intuitive picture, one is to expect:
i)~negative exchange bias ($H_E < 0$); ii)~that uncompensated
interfaces should display the largest magnitudes of $|H_E|$; and,
iii)~that roughness of a compensated interface should increase
$|H_E|$. Even a cursory inspection of the experimental
results~\cite{nogu99_rev} shows that {\bf none} of these expectations
is fulfilled.

\subsection{The ground breaking contribution of N\'eel}
\label{neel_theory}

Ten years after the discovery of EB, N\'eel~\cite{neel67} formulated a
model that applies to a system which consists in a weakly anisotropic
uncompensated AF interface layer (see Fig.~\ref{fig:magn_colin})
ferromagnetically coupled across the interface to a F slab. He
assumed that the magnetization $\vec m_i$ of layer $i$, both in the F
and in the AF, is uniform within the layer and parallel to the
interface.  Adopting as the unit of length the lattice parameter $a =
1$, the condition for $\vec m_i$ to be in equilibrium is

\begin{equation}
JS^2 \ [\sin \frac 12 (\theta_{i+1} - \theta_{i}) +
\sin \frac 12 (\theta_{i-1} - \theta_{i})] 
-2K \sin \theta_i = 0 \; ,
\label{eq:neel1}
\end{equation}

\noindent where $\frac 12 \theta_i$ is the angle between $\vec m_i$
and the easy magnetization axis, and $J$ and $K$ were defined after
Eq.~\ref{eq:meik}. In the continuum approximation the above set of
difference equations becomes the following differential equation: 

\begin{equation}
  JS^2 \frac {d^2 \theta}{di^2} - 4K \sin \theta = 0 \; .
\label{eq:neel2}
\end{equation}

Solving the above equation for specific values of $J$ and $K$, with
the assumption of uniaxial anisotropy, N\'eel was able to derive the
magnetization profile. Under appropriate conditions domains develop
both in the F and in the AF, but the continuum approximation requires
a minimum width of the F and AF slabs to be valid; for example, a
ferromagnetic iron slab in excess of 1000~\AA~ is needed.  Thus,
while the N\'eel model is an important milestone, its application to
the better characterized and well-controlled thin film EB systems
developed recently is quite restricted and has to be implemented with
due caution. 

\subsection{Early random interface model}
\label{maloz}

\begin{figure}
\begin{center}
\includegraphics[height=21cm]{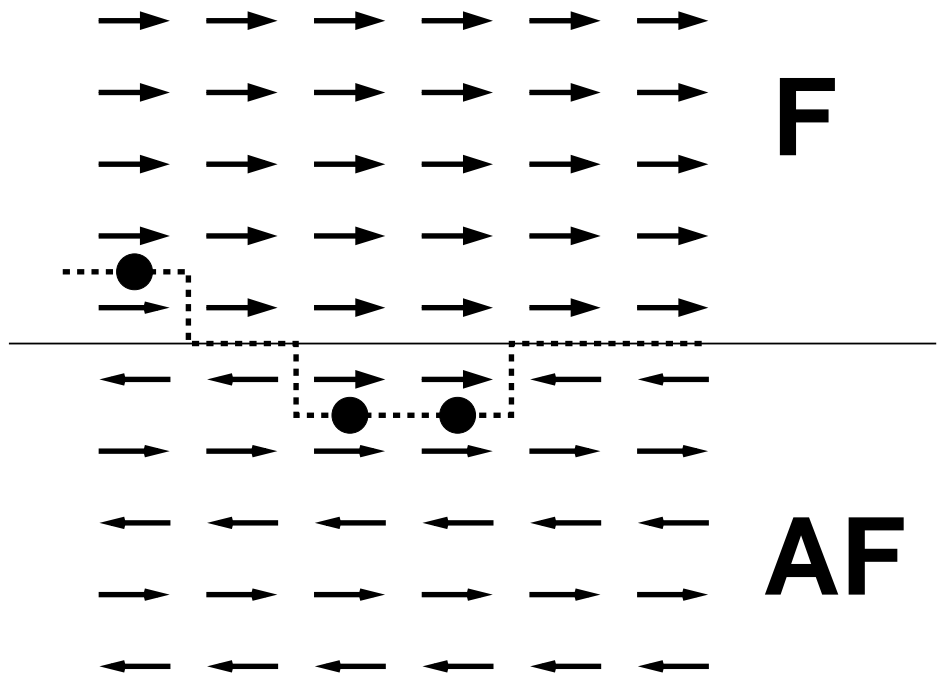}
\caption{AF coupled rough interface with frustrated interactions
  marked by full dots . The dashed line marks the boundary between the
  F and the AF.}
\label{fig:maloz}
\end{center}
\end{figure}

Twenty years after N\'eel's publication Malozemoff~\cite{malo87}, in
1987, proposed a model of exchange anisotropy based on the assumption
of rough F/AF compensated and uncompensated interfaces, as
illustrated in Fig.~\ref{fig:maloz}. Random interface roughness gives
rise to a random magnetic field that acts on the interface spins,
yielding unidirectional anisotropy. The latter causes the asymmetric
offset of the hysteresis loop. This way it is possible to reconcile
the experimental data with theory, reducing by two orders of
magnitude the overestimate derived using Eq.~\ref{eq:overestimate}.
The expression given in Ref.~\cite{malo87}, for the shift ${H}_E$ of
the hysteresis loop, is

\begin{equation}
  \text{H}_E = \frac {2}{M_F t_F} {\sqrt \frac {J_{AF} K_{AF}}{a}} \; ,
\label{eq:maloz1}
\end{equation}

\noindent where I use the same notation as in
Eq.~\ref{eq:overestimate}. The reduction factor of the original
estimate (ignoring the differences between the various exchange and
anisotropy constants) is $2 a / \sqrt{J / a K}$, which corresponds to
the ratio of twice the lattice parameter $a$ divided by the F domain
wall~\cite{kittel_ISSP} width $d_w$, since $d_w \sim \sqrt{J / a K}$.
The role played by the ratio $a / d_w$ underscores the fact that the
characteristic length scale of this problem is $d_w$. A refinement of
the above ideas, put forward in the same paper~\cite{malo87}, reduces
the ratio still further by allowing for the formation of AF domain
walls in the vicinity of the interface (the estimate for $d_w$ is
assumed to hold both for the F and AF).

In spite of its success in obtaining a reasonable estimate for H$_E$,
this model has a severe drawback: it crucially depends on a defect
concentration at the interface which is not consistent with
experiments, as will be discussed in detail further on in this
review.  However, it was recently reexamined and extended by
Schulthess and Butler~\cite{schu98,schu99}, as addressed below in
Sec.~\ref{random}.

\subsection{AF domain wall models}
\label{mauri}

Shortly after Malozemoff's proposal an alternative suggestion was
advanced by Mauri {\it et al.}~\cite{mauri87} (while usually referred
as ``the Mauri model'', it is coauthored by Mauri, Siegmann, Bagus and
Kay). The main assumptions made are: i)~F interface coupling across a
perfect flat interface; ii)~parallel magnetization of the F and AF
sublattices in the absence of an external field; iii)~a F slab
thickness $t_F$ much smaller than the F domain wall width; and iv)~a
domain wall (DW) that develops inside the AF, which has the effect of
imposing an upper limit on the exchange coupling energy, such that it
reaches significantly smaller values than those given by
Eq.~\ref{eq:overestimate}. Assumptions i), ii) and iv) are debatable;
first, AF interface coupling is not only possible but most likely in
several cases. In fact Nogu\'es {\it et al.}~\cite{nogu00+} have
confirmed experimentally that AF interface coupling is necessary to
observe positive exchange bias.  Moreover, it does not provide clues
to understand how compensated interfaces can yield values of H$_E$ as
large, or even larger, than uncompensated ones~\cite{nogu99_rev}.
Furthermore, in the magnetic ground state configuration, the F magnetic
moments are orthogonal to the bulk AF easy axis (as pointed out by
Koon~\cite{koon97} and confirmed experimentally by Moran {\it et
  al.}~\cite{mora98} and Ijiri~{\it et al.}~\cite{ijir98}).  Finally,
for a DW to develop in the AF, the anisotropy constant $K_{AF}$ has to
be quite small; otherwise it is energetically favorable for the DW to
form in the F side, as inferred experimentally in
Refs.~\cite{ball96,fitz00,mill96,stro97,nogu99b,full98} and argued
theoretically in Refs.~\cite{kmpr1,kmpr2,kmpr3}.

\subsection{Orthogonal F and AF magnetic lattices} 
\label{koon}

In 1997 Koon~\cite{koon97} tackled the problem of explaining EB in
thin films with compensated F/AF interfaces by means of a
micromagnetic calculation. His main result was to establish, on the
basis of a Heisenberg model, that the ground state configuration
corresponds to {\it perpendicular} orientation of the bulk F moments
relative to the AF magnetic easy axes direction, as illustrated in
Fig.~\ref{fig:koon}. Moreover, Koon also showed that the magnetic
moments in the AF interface layer exhibit canting; in fact, the
minimum energy is achieved with the AF spins adopting a relatively
small canting angle ($\theta < 10^\circ$) relative to the AF bulk
easy axis, with a component opposite to the cooling field direction.

While the work of Koon is relevant in establishing the right
interface magnetic structure unfortunately, as properly pointed out
by Schulthess and Butler~\cite{schu98}, it fails to yield EB. In
other words, the canted interface magnetic structure by itself is not
sufficient to generate EB, {\it i.e.} to produce the required {\it
unidirectional} anisotropy and the consequent shifted hysteresis
loops with H$_E \neq 0$.

\subsection{Random interface field models} 
\label{random}

Schulthess and Butler~\cite{schu98,schu99} showed that Malozemoff's
random interface field and Koon's orthogonal magnetic arrangement,
rather than being in conflict, could be combined to provide an
explanation of EB. In their model they added to the usual exchange,
Zeeman and anisotropy energies, the dipolar interaction term $E_D$

\begin{equation}
  E_D = \sum_{i\neq j} \frac {[\vec \mu_i \cdot \vec \mu_j 
  - 3 (\vec  \mu_i \cdot \hat n_{ij}) 
  (\vec \mu_j \cdot \hat n_{ij}) ]} 
  {|\vec R_i -  \vec R_j|^3} \; ,
\label{eq:schult}  
\end{equation}

\noindent where $\{\mu_i\}$ is the magnetic moment configuration and
$\hat n_{ij}$ is a unit vector parallel to $\vec R_i - \vec R_j$.
Magnetic properties were obtained using a classical micromagnetic
approach~\cite{labo69,brow78,ahar96}, solving the Landau-Lifshitz
equations of motion, including a Gilbert-Kelley damping term, in
order to attain stable or metastable equilibrium.

As already mentioned, when the above model is applied to the Koon
orthogonal interface configuration, illustrated in
Fig.~\ref{fig:koon}, for flat interfaces the coupling that results
does not yield unidirectional anisotropy, but rather irreversible
magnetization curves with finite coercivity. The irreversibility
appears as a bifurcation in the solution of the equation of motion.
Thus, additional elements are required to generate exchange bias.
Following the spirit of Malozemoff's model, in
Refs.~\cite{schu98,schu99} surface defects were introduced by
assuming a 4$\times$4 2D interface unit cell, with one interfacial F
site occupied by an AF magnetic moment.  This way values of H$_E$,
and of the coercivity H$_c$, of comparable magnitude to experimental
observation~\cite{taka97} for the CoO/F system (F: Co and permalloy)
are obtained, when exchange and anisotropy parameters of reasonable
magnitude, and a canting angle of 10$^\circ$, are adopted. Of course
there is a caveat: the model hinges qualitatively on the assumption
of a rough interface, and the quantitative results depend on the
nature and concentration of the interface defects that are
incorporated.

In this context it is pertinent to stress that the relation between
surface roughness and EB is quite complex experimentally, and far
from understood theoretically.  Experimentally Moran {\it et
al.}~\cite{mora95} established already in 1995 that interface
disorder {\it increases} H$_E$ in the permalloy/CoO system. Very
recently Leighton {\it et al.}~\cite{leig00} reported an even more
surprising result: that as a function of interface roughness both the
coercivity H$_c$ and H$_E$ can manifest quite unexpected behaviors.
For example, in the Fe/FeF$_2$ system the rougher the interface the
larger H$_E$, but the opposite occurs in the very similar Fe/MnF$_2$
system. Moreover, Fe/MnF$_2$ exhibits large changes of both H$_E$ and
H$_c$, as a function of the cooling field strength H$_{cf}$, when the
interface is smooth, but hardly a significant variation when it is
rough.

Also Zhang {\it et al.}~\cite{zhan99} investigated theoretically the
coercivity of EB systems induced by random fields at the F/AF
interface.  They incorporated domain walls on the F side of the
interface and derived the correct order of magnitude and temperature
dependence of the coercivity (H$_c \propto T^{-3/2}$). Still another
random field model was investigated by Dimitrov {\it et
al.}~\cite{dimi98}, starting from the assumption of an interface
exchange interaction between F and AF magnetic moments $\vec {\bf
m}_{\text{F}}$ and $\vec {\bf m}_{\text{AF}}$, respectively, given by 

\begin{equation}
  \varepsilon = J_1 \vec {\bf m}_{\text{F}} \cdot 
                \vec {\bf m}_{\text{AF}}
                + J_2 (\vec {\bf m}_{\text{F}} \cdot 
                \vec {\bf m}_{\text{AF}})^2  \; ,
\label{eq:dimitrov1}
\end{equation}

\noindent where $J_1$ and $J_2$ are the normal and
biquadratic~\cite{slon91} exchange constants.  $J_1$ favors parallel
or antiparallel alignment, while $J_2$ favors orthogonal (spin flop
like) F/AF coupling. Summing over all the interactions the above
expression leads to the following form for the total energy:

\begin{equation}
  E = C_1 + C_2 J_1 cos \theta + C_3 J_2 sin^2  \theta \; ,
\label{eq:dimitrov2} 
\end{equation}

\noindent where $\theta $ is the angle formed by the easy axes of 
the F and AF and the $C_k$'s are coefficients which cannot be
calculated without detailed interface information. Using an educated
guess for the pertinent parameters several qualitatively correct
conclusions were obtained from this model.

\subsection{The frozen interface model} 
\label{kmpr}

\begin{figure}
\begin{center}
\includegraphics[height=20cm]{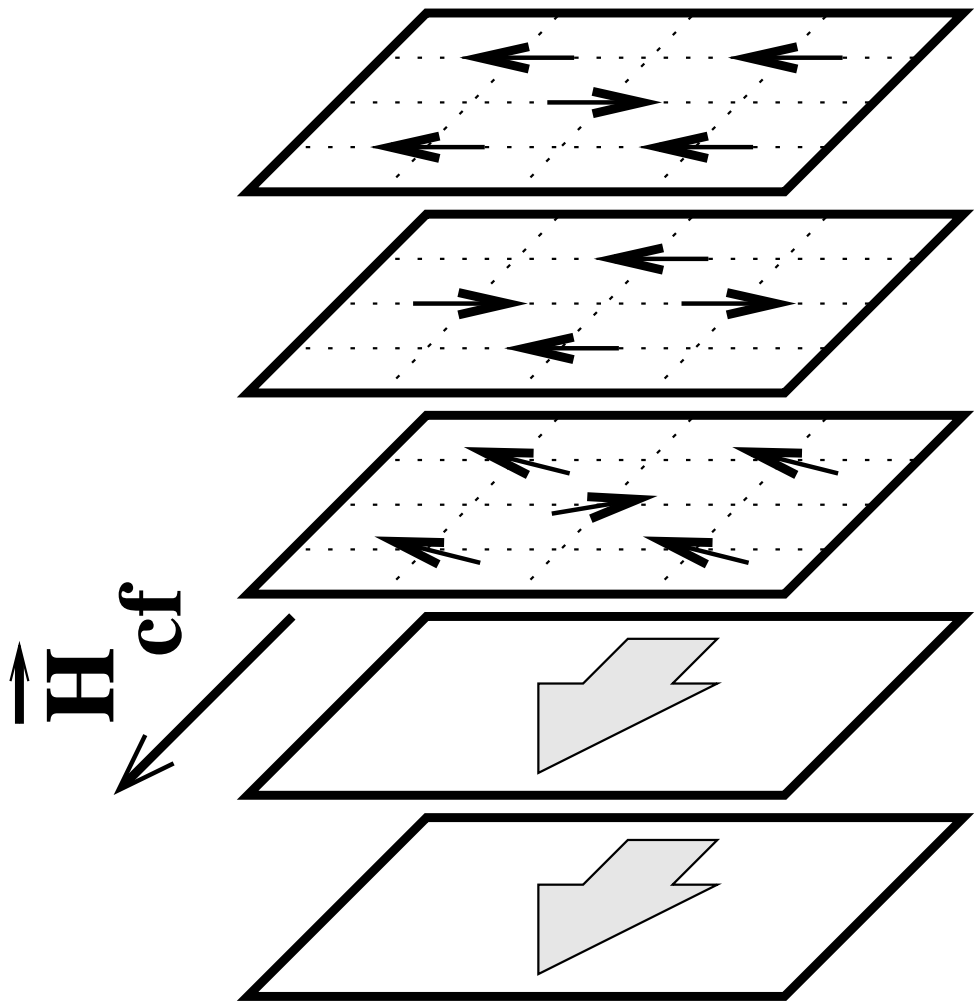}
 \caption{Illustration of the perpendicular F and AF
magnetic interface configuration, with spin canting in the first AF
layer.}  
\label{fig:koon}
\end{center}
\end{figure}

Recently Kiwi {\it et al.}~\cite{kmpr1,kmpr2,kmpr3} put forward an EB
model which applies to a large variety of systems where the
anisotropy of the AF is relatively large, and thus the energy cost of
creating a domain wall in the AF quite considerable.  In particular
Fe/FeF$_2$ and Fe/MnF$_2$ were adopted as prototypes, since there is
extensive experimental information on them, they have precisely
characterized interfaces, a rather simple crystal and interface
structure, and large H$_E$
values~\cite{nogu99_rev,nogu96,leig00,nogu99,nogu00}. Attention was
focused on the (110) {\it compensated} AF crystal face, which
exhibits the largest EB. The zero applied field interface spin
configuration is described by the illustrative cartoon provided as
Fig.~\ref{fig:koon}.  This spin configuration is a consequence of the
fact that the two characteristic length scales in the problem (the F
and AF domain wall widths, $d_w^F$ and $d_w^{AF}$, respectively) are
very different~\cite{stil99}.  While $d_w^{Fe} \sim$ 100 nm, $\
d_w^{AF}$ amounts to just a few monolayers due to the large FeF$_2$
anisotropy, which is consistent with results by Carri\c co {\it et
al.}~\cite{carr94}.

As in all models an assumption is made: that the first AF interface
layer freezes into the canted spin configuration it adopts close to
$T_N$. Since $t < d_w^F$, where $t$ is the thickness of the F slab, a
discrete treatment is in order.  Analytically

\begin{equation}
{\cal{H}} = {\cal{H}}_{AF} + {\cal{H}}_{F/AF} + {\cal{H}}_F \; ,
\label {Ham}
\end{equation}

\noindent where ${\cal{H}}_{AF}$, ${\cal{H}}_{F/AF}$ and
${\cal{H}}_{F}$ describe the AF substrate, interface coupling and the
F slab, respectively. For the single magnetic cell, partially
represented in Fig.~\ref{fig:koon}. These terms can be written as

\begin{eqnarray}
{\cal{H}}_{AF} = &-& J_{AF} \; [ \; S\; \hat e_{AF}  \cdot 
(\vec S^{(\alpha)} - \vec S^{(\beta)}) 
+ 2  \vec S^{(\alpha)} \cdot \vec S^{(\beta)} \; ] 
\label {Ham_AF}\\
&-& \frac 12 K_{AF}\; [\;  (\vec S^{(\alpha)} \cdot  \hat e_{AF})^2\; 
+ \; (\vec S^{(\beta)}  \cdot  \hat e_{AF})^2\;  ] 
- \frac 12 \mu_B g \; (\vec S^{(\alpha)} + \vec S^{(\beta)}) 
\cdot \vec H \; , \nonumber \\ 
{\cal{H}}_{F/AF} = &-& J_{F/AF} \; (\vec S^{(\alpha)} + \vec S^{(\beta)}) 
\cdot \vec S_1 \; , 
\label {Ham_F/AF}\\
{\cal{H}}_{F} = &-&  2 J_{F} \; \sum_{k=1}^{N-1}
\vec S_k \cdot \vec S_{k+1}
- \sum_{k=1}^{N} [\; \frac {K_{F}}{H^2} \; (\vec S_k \cdot \vec H)^2  
+ \mu_B g \; \vec S_k \cdot \vec H\; ]\; .
\label {Ham_F}
\end{eqnarray}

\noindent Above $S = |\vec S|$ and $N$ is the number of F layers.
$\mu_B$ and $g$ denote the Bohr magneton and the Fe gyromagnetic
ratio, respectively, while $\vec H$ is the external applied magnetic
field.  $J_{\mu}$ denotes the Heisenberg exchange parameter and
$K_{\mu}$ the uniaxial anisotropy. In Eq.~\ref{Ham_AF} the unit
vector $\hat e_{AF}$ defines the AF uniaxial anisotropy direction,
$\vec S^{(\alpha)}$ and $\vec S^{(\beta)}$ are canted spin vectors in
the AF interface, belonging to the $\alpha$- and $\beta$-AF
sub-lattices.  The vectors $\vec S_k$ are the spin vectors of the
$k$-th F layer, with $k=1$ labels the F interface, $1\leq k \leq N$,
where the value $N=65$ was adopted for the total number of F layers.
For Fe this corresponds to $t_F\approx$ 13 nm, where $t_F$ denotes
the width of the F slab.
 
Labeling $\theta^{(\alpha)} \ (\theta^{(\beta)})$ as the average angle
between $\vec S^{(\alpha)}$ ($\ \vec S^{(\beta)}$) and the cooling
field $\vec H_{cf}$ and assuming $\theta^{(\alpha)} =
-\theta^{(\beta)}$, the set of nonlinear equations to be solved for
$\{\theta_k\}$ is

\begin{eqnarray}
&h& \sin \theta_j 
- (1 - \delta_{j,N}) \sin (\theta_{j+1} - \theta_{j}) 
+ \delta_{j,1} \ \kappa \sin \theta_1 \nonumber \\ 
&+& (1 - \delta_{j,1})  \sin  (\theta_j - \theta_{j-1})
+ 2 D \sin \theta_j \cos \theta_j = 0 \; ,
\label {deriv}
\end{eqnarray}

\noindent where $\delta_{i,j}$ the Kronecker symbol, $h =
\mu_B g H / 2J_F < 10^{-3}$, $\mu_B$ is the Bohr magneton, $D = K_F /
2 J_F< 10^{-5}$, $\kappa$ is the effective interface coupling, and $J$
and $K$ denote the exchange and anisotropy parameters, respectively.

This set of equations can be solved using Camley's
method~\cite{caml87,caml88} and by simulated annealing~\cite{kirk83}.
Both yield the same values of H$_E$ which, using a single adjustable
parameter, the interface coupling constant $J_{F/AF}$, agree with
experiment~\cite{kmpr1}. Moreover, the calculations show that
H$_E\propto t_F^{-1}$, as long as the F thickness $t_F < d_w^F$. The
energy is reversibly stored in an incomplete domain wall, or magnetic
like structure~\cite{full98b}, in the F slab. The magnetic
structure of this incomplete domain wall in the F has a twist smaller
than 20$^\circ$, and is qualitatively compatible with the neutron
scattering experiment results obtained by Ball {\it et
al.}~\cite{ball96}.

However, the strongest experimental support for the above picture was
obtained by very recent experiments. Nolting {\it et
al.}~\cite{nolt00} established that the alignment of the spins in
individual F domains close to the interface is determined, domain by
domain, by the spin direction in the underlying AF. Even more
detailed support is provided by the scanning electron microscopy
imaging experiments of Matsuyama {\it et al.}~\cite{mats00}. They
investigated Fe domains deposited on the fully compensated (001) face
of NiO and observed that the Fe spin polarization of each domain is
roughly perpendicular to an easy axis of the NiO. Moreover, they also
infer that the NiO interface spins cant in relation to the Fe spins.

The model by Kiwi {\it et al.} also allows for a simple explanation
of positive exchange bias~\cite{kmpr3}, which is in fairly good
agreement with experiment. The positive exchange bias problem had
been addressed previously by Hong~\cite{hong98}, on the basis of a
spin wave theory put forward by Suhl and Schuller~\cite{suhl98}. In
the latter approach~\cite{suhl98} the mechanism that generates EB is,
to the best of my knowledge, the only one that does not introduce
{\it ad hoc} assumptions about the interface structure, since the
coupling is a consequence of the emission and reabsorption, by a
ferromagnetic spin, of virtual AF spin waves across the interface.
According to Hong~\cite{hong98} a strong cooling field polarizes the
AF spins in the opposite direction to the low field cooled ones,
which results in H$_E > 0$.

\subsection{Local pinning field variation} 
\label{stiles}

Stiles and McMichael~\cite{stil99} adopted a conceptually different
approach. Rather than focusing their attention on the interaction of
individual atoms, or magnetic moments, they constructed their theory
using interface AF grains, with stable magnetic order, as building
blocks of a polycrystalline interface. They assume that these
AF magnetic grains, which in the absence of the F slab can order in
many different quasi-degenerate arrangements, choose a particular
stable energy configuration when in contact with the F.  Because of
the weakness of the Zeeman term, this interface magnetic configuration
is stable and retains a ``memory'' of the initial F direction, {\it
  i.e.} the direction of the F magnetization when AF order sets in.
Moreover, they suggest that due to the polycrystalline nature of the
system under scrutiny, even for uncompensated AF interfaces, there is
a substantial compensation of the magnetic moments due to the
fluctuating easy axis direction of each individual grain. Thus, in
this model~\cite{stil99}, a fraction of uncompensated spins at the
interface drives the unidirectional anisotropy.

The starting point of the calculation is to consider a single domain
AF grain. The energy of each grain that is coupled to the F is given
by 

\begin{equation}
  \frac E {Na^2} = \frac {-J_{net}}{a^2} [\hat {\bf M}_{\text{FM}}\cdot 
    \hat {\bf m}(0)]
    +   \frac {J_{sf}}{a^2} [\hat {\bf M}_{\text{FM}}\cdot 
    \hat {\bf m}(0)]^2 
    + \frac 12 \sigma [1 - \hat {\bf m}(0) \cdot ({\bf \pm \hat u})] \; ,
\label{eq:stiles}
\end{equation}

\noindent where $a$ is the lattice constant and $\hat {\bf
  M}_{\text{FM}}$, $\hat {\bf m}(0)$ and ${\bf \pm \hat u}$ establish
the F magnetization, net AF sublattice magnetization and the two easy
uniaxial anisotropy directions, respectively. $J_{net}$ is the average
coupling energy to the net moment of the AF grain, $J_{sf}$ is the
spin flop energy and $\sigma$ is the energy of a 180$^\circ$ domain
wall in the AF. Thus, there is a competition between parallel
alignment described by $J_{net}$ and perpendicular, spin flop-like,
alignment induced by $J_{sf}$. The formal similarity between
Eq.~\ref{eq:stiles} and Eq.~\ref{eq:dimitrov1} is quite obvious;
however, it has to be stressed that the latter deals with single
magnetic moments while the former applies to AF interface grains.
Moreover, the possibility of a domain wall forming in the AF is
additionally incorporated and described by the term proportional to
$\sigma$.

On the basis of the above outlined model they calculate the relevant
physical properties of the system, {\it i.e.} the magnitude of the
unidirectional anisotropy and the hysteretic effects that induce
coercivity, as well as the consequences for field rotational torque
and ferromagnetic resonance measurements.  This is done for a variety
of parameter values, both ignoring and including spin flop-like
coupling. 

While a satisfactory description is thus achieved, several assumptions
have to be made, in addition to those outlined above in relation to
the justification of Eq.~\ref{eq:stiles}. To lock the interface spin
configuration, partial domain walls are required to wind up in the
AF. Moreover, it is postulated that for some AF grains a critical
winding angle exists which, if exceeded, leads to instability of the
AF order. This way the AF grains can either support a particular AF
order or they can switch between two possible states; the former is
associated with reversible, and the latter with hysteretic, behavior.

\section{Summary and discussion}
\label{final}

The challenge posed by the EB phenomenon has generated the vigorous
activity described in the preceding sections. It is apparent that the
effort to probe the EB phenomenon experimentally, and to understand
it theoretically, has truly flourished during the last years.  As is
often the case, the simple systems, that are more amenable to be
grasped by theory ({\it e.g.} atomically ordered epitaxially grown
bilayers), are less relevant for technological applications, since
polycrystalline thin film multilayers are the ones routinely employed
in actual devices.  However, the insight derived from the
understanding of simple systems may well prove transferable to more
complex cases.

As discussed in Sec.~\ref{interf_str}, at present the major obstacle in
the path to a full understanding of EB is the knowledge of the
crystallographic and magnetic structure in the vicinity of the F/AF
interface. Thus, interface sensitive experimental probes, like the
recently published work of Nolting {\it et al.}~\cite{nolt00} and
Matsuyama {\it et al.}~\cite{mats00}, as well as the grazing angle
neutron scattering experiments by Ball {\it et al.}~\cite{ball96} and
Fitzsimmons {\it et al.}~\cite{fitz00}, provide important clues to
improving our understanding of the phenomenon. 

Besides the above uncertainties, theorists also have to deal properly
with the intrinsically complex and subtle mechanisms involved in
EB. For example, it took more than four decades from the discovery of
EB by Meiklejohn and Bean~\cite{meik56} to the realization by
Koon~\cite{koon97} that the F bulk magnetization is orthogonal to the
AF sublattice magnetic moments. To order zero (perfectly ordered
magnetic and crystal bulk structures all the way up to the interface)
the collinear and orthogonal configurations require the same energy.
In other words, the energy cost of the collinear configuration (one
half of the interface magnetic moments frustrated) is the same as for
perpendicular ordering (all the interface moments half-frustrated).
It is only when canting in the vicinity of the interface is
incorporated in the calculation that the perpendicular ordering
proves to be more favorable. Thus the importance of the contribution
by Koon, in spite of the fact that the model~\cite{koon97} fails to
yield H$_E \neq 0$. It is worth mentioning that the orthogonal
configuration was confirmed experimentally by Moran {\it et
al.}~\cite{mora98} and Ijiri~{\it et al.}~\cite{ijir98}.

In addition, as outlined in Sec.~\ref{introduction}, there are many
systems that exhibit EB and it is quite likely that no single theory
will be able to properly fit and describe all of them. In fact,
it might well be that each, or at least some, of the assumptions
advanced on the interface structure apply to different classes of
systems. The relative strength of the anisotropy parameters $K_F$,
$K_{F/AF}$ and $K_{AF}$ also is an important physical quantity in
determining which theory is applicable to which system.

Thus, at this point it seems adequate to make a critical evaluation
of where we stand, in terms of the list of requirements specified in
Sec.~\ref{introduction}, for a sound theory of EB. I will go
through them point by point: 1)~As far as the essential issue of
establishing a mechanism {\it free of ad hoc assumptions on the
interface structure} that yields unidirectional anisotropy, it is clear
that there is a long way to go before that goal is achieved; 2)~based
on the several models put forward for the crystallographic and
magnetic structure of the interface reasonable values of H$_E$ and
H$_c$ have been obtained.  Also, the dependence of H$_E$ on H$_{cf}$
has been derived, but less so the temperature dependence of these
quantities~\cite{neel67,neel88,mauri87,malo87,malo88a,malo88b,schu98,schu99,stil99,kmpr1,kmpr2,kmpr3,taka97,zhan99,dimi98};
3)~on the basis of the orthogonal interface spin
arrangement the rationale for the large H$_E$ values that compensated
AF interfaces exhibit are now on reasonably firm
ground~\cite{schu98,schu99,stil99,kmpr1,kmpr2,kmpr3};
4,5,6 and 7)~on the contrary, the relation between interface roughness
and EB remains a mystery to theory, as well as the memory effect and
its relation to the blocking temperature, the training effect and the
asymmetry of the hysteresis loop.

In conclusion, the abundant new experimental information and the
refined theories put forward during the last five years have allowed
investigators to make significant headway in the description,
understanding and technological use of the exchange bias phenomenon,
but it is equally clear that many important issues remain open.

\acknowledgments 

I thank Professors Ivan K.  Schuller, David Lederman and Josep
Nogu{\'e}s for valuable suggestions and the critical reading of the
manuscript.  Stimulating discussions with Jos\'e Mej\'{\i}a-L\'opez,
Ruben Portugal and Ricardo Ram\'{\i}rez are gratefully acknowledged.
This work was supported by the Fondo Nacional de Investigaciones
Cient\'{\i}ficas y Tecnol\'ogicas (FONDECYT, Chile) under grant
\#~8990005.

\newpage
\bibliographystyle{apsrev}
\bibliography{EBbibl}

\begin{thebibliography}{10}
\expandafter\ifx\csname bibnamefont\endcsname\relax
  \def\bibnamefont#1{#1}\fi
\expandafter\ifx\csname bibfnamefont\endcsname\relax
  \def\bibfnamefont#1{#1}\fi
\expandafter\ifx\csname url\endcsname\relax
  \def\url#1{\texttt{#1}}\fi
\expandafter\ifx\csname urlprefix\endcsname\relax\def\urlprefix{URL }\fi
\expandafter\ifx\csname bibinfo\endcsname\relax \def\bibinfo#1#2{#2}\fi
\expandafter\ifx\csname eprint\endcsname\relax \def\eprint#1{#1}\fi

\bibitem{meik56}
\bibinfo{author}{\bibfnamefont{W.~P.} \bibnamefont{Meiklejohn}}
  \bibnamefont{and} \bibinfo{author}{\bibfnamefont{C.~P.} \bibnamefont{Bean}},
  \bibinfo{journal}{Phys. Rev.} \textbf{\bibinfo{volume}{102}},
  \bibinfo{pages}{1413} (\bibinfo{year}{1956}).

\bibitem{nogu99_rev}
\bibinfo{author}{\bibfnamefont{J.}~\bibnamefont{Nogu{\'e}s}} \bibnamefont{and}
  \bibinfo{author}{\bibfnamefont{I.~K.} \bibnamefont{Schuller}},
  \bibinfo{journal}{J. Magn. Magn. Mat.} \textbf{\bibinfo{volume}{192}},
  \bibinfo{pages}{203} (\bibinfo{year}{1999}), \bibinfo{note}{and references
  therein.}

\bibitem{berk99}
\bibinfo{author}{\bibfnamefont{A.~E.} \bibnamefont{Berkowitz}}
  \bibnamefont{and} \bibinfo{author}{\bibfnamefont{K.}~\bibnamefont{Takano}},
  \bibinfo{journal}{J. Magn. Magn. Mater.} \textbf{\bibinfo{volume}{200}},
  \bibinfo{pages}{552} (\bibinfo{year}{1999}), \bibinfo{note}{and references
  therein.}

\bibitem{stam00}
\bibinfo{author}{\bibfnamefont{R.~L.} \bibnamefont{Stamps}},
  \bibinfo{journal}{J. Phys. D: Appl. Phys.} \textbf{\bibinfo{volume}{33}},
  \bibinfo{pages}{R247} (\bibinfo{year}{2000}), \bibinfo{note}{and references
  therein.}

\bibitem{nogu96}
\bibinfo{author}{\bibfnamefont{J.}~\bibnamefont{Nogu{\'e}s}},
  \bibinfo{author}{\bibfnamefont{D.}~\bibnamefont{Lederman}},
  \bibinfo{author}{\bibfnamefont{T.~J.} \bibnamefont{Moran}}, \bibnamefont{and}
  \bibinfo{author}{\bibfnamefont{I.~K.} \bibnamefont{Schuller}},
  \bibinfo{journal}{Phys. Rev. Lett.} \textbf{\bibinfo{volume}{76}},
  \bibinfo{pages}{4624} (\bibinfo{year}{1996}).

\bibitem{shen96}
\bibinfo{author}{\bibfnamefont{J.~X.} \bibnamefont{Shen}} \bibnamefont{and}
  \bibinfo{author}{\bibfnamefont{M.~T.} \bibnamefont{Kief}},
  \bibinfo{journal}{J. Appl. Phys.} \textbf{\bibinfo{volume}{79}},
  \bibinfo{pages}{5008} (\bibinfo{year}{1996}).

\bibitem{char71}
\bibinfo{author}{\bibfnamefont{S.~H.} \bibnamefont{Charp}} \bibnamefont{and}
  \bibinfo{author}{\bibfnamefont{E.}~\bibnamefont{Fulcomer}},
  \bibinfo{journal}{J. Appl. Phys.} \textbf{\bibinfo{volume}{42}},
  \bibinfo{pages}{1426} (\bibinfo{year}{1971}).

\bibitem{fulc72}
\bibinfo{author}{\bibfnamefont{E.}~\bibnamefont{Fulcomer}} \bibnamefont{and}
  \bibinfo{author}{\bibfnamefont{S.~H.} \bibnamefont{Charp}},
  \bibinfo{journal}{J. Appl. Phys.} \textbf{\bibinfo{volume}{43}},
  \bibinfo{pages}{4184} (\bibinfo{year}{1972}).

\bibitem{tsan82}
\bibinfo{author}{\bibfnamefont{C.}~\bibnamefont{Tsang}} \bibnamefont{and}
  \bibinfo{author}{\bibfnamefont{K.}~\bibnamefont{Lee}}, \bibinfo{journal}{J.
  Appl. Phys.} \textbf{\bibinfo{volume}{53}}, \bibinfo{pages}{2605}
  (\bibinfo{year}{1982}).

\bibitem{ambr97}
\bibinfo{author}{\bibfnamefont{T.}~\bibnamefont{Ambrose}},
  \bibinfo{author}{\bibfnamefont{R.~L.} \bibnamefont{Sommer}},
  \bibnamefont{and} \bibinfo{author}{\bibfnamefont{C.~L.} \bibnamefont{Chien}},
  \bibinfo{journal}{Phys. Rev. B} \textbf{\bibinfo{volume}{56}},
  \bibinfo{pages}{83} (\bibinfo{year}{1997}).

\bibitem{chou98}
\bibinfo{author}{\bibfnamefont{S.~M.} \bibnamefont{Chou}},
  \bibinfo{author}{\bibfnamefont{K.}~\bibnamefont{Liu}}, \bibnamefont{and}
  \bibinfo{author}{\bibfnamefont{C.~L.} \bibnamefont{Chien}},
  \bibinfo{journal}{Phys. Rev. B} \textbf{\bibinfo{volume}{58}},
  \bibinfo{pages}{R14717} (\bibinfo{year}{1998}).

\bibitem{wu98}
\bibinfo{author}{\bibfnamefont{X.~W.} \bibnamefont{Wu}} \bibnamefont{and}
  \bibinfo{author}{\bibfnamefont{C.~L.} \bibnamefont{Chien}},
  \bibinfo{journal}{Phys. Rev. Lett.} \textbf{\bibinfo{volume}{2795}},
  \bibinfo{pages}{2795} (\bibinfo{year}{1998}).

\bibitem{goke99}
\bibinfo{author}{\bibfnamefont{N.~J.} \bibnamefont{G{\"o}kemeijer}}
  \bibnamefont{and} \bibinfo{author}{\bibfnamefont{C.~L.} \bibnamefont{Chien}},
  \bibinfo{journal}{J. Appl. Phys.} \textbf{\bibinfo{volume}{85}},
  \bibinfo{pages}{5516} (\bibinfo{year}{1999}).

\bibitem{goke99b}
\bibinfo{author}{\bibfnamefont{N.~J.} \bibnamefont{G{\"o}kemeijer}},
  \bibinfo{author}{\bibfnamefont{J.~W.} \bibnamefont{Cai}}, \bibnamefont{and}
  \bibinfo{author}{\bibfnamefont{C.~L.} \bibnamefont{Chien}},
  \bibinfo{journal}{Phys. Rev. B} \textbf{\bibinfo{volume}{60}},
  \bibinfo{pages}{3033} (\bibinfo{year}{1999}).

\bibitem{li00}
\bibinfo{author}{\bibfnamefont{Y.~F.} \bibnamefont{Li}},
  \bibinfo{author}{\bibfnamefont{R.~H.} \bibnamefont{Yu}},
  \bibinfo{author}{\bibfnamefont{J.~Q.} \bibnamefont{Xiao}}, \bibnamefont{and}
  \bibinfo{author}{\bibfnamefont{D.~V.} \bibnamefont{Dimitrov}},
  \bibinfo{journal}{J. Appl. Phys.} \textbf{\bibinfo{volume}{87}},
  \bibinfo{pages}{4951} (\bibinfo{year}{2000}).

\bibitem{ball96}
\bibinfo{author}{\bibfnamefont{A.~R.} \bibnamefont{Ball}},
  \bibinfo{author}{\bibfnamefont{A.~J.~G.} \bibnamefont{Leenaers}},
  \bibinfo{author}{\bibfnamefont{P.~J.} \bibnamefont{van~der Zaag}},
  \bibinfo{author}{\bibfnamefont{K.}~\bibnamefont{Shaw}},
  \bibinfo{author}{\bibfnamefont{B.}~\bibnamefont{Singer}},
  \bibinfo{author}{\bibfnamefont{D.~M.} \bibnamefont{Lind}},
  \bibinfo{author}{\bibfnamefont{H.}~\bibnamefont{Fredrikze}},
  \bibnamefont{and} \bibinfo{author}{\bibfnamefont{M.~T.}
  \bibnamefont{Rekveldt}}, \bibinfo{journal}{Appl. Phys. Lett.}
  \textbf{\bibinfo{volume}{69}}, \bibinfo{pages}{1489} (\bibinfo{year}{1996}).

\bibitem{leig00}
\bibinfo{author}{\bibfnamefont{C.}~\bibnamefont{Leighton}},
  \bibinfo{author}{\bibfnamefont{J.}~\bibnamefont{Nogu{\'e}s}},
  \bibinfo{author}{\bibfnamefont{B.~J.} \bibnamefont{{J\"onsson-\AA kerman}}},
  \bibnamefont{and} \bibinfo{author}{\bibfnamefont{I.~K.}
  \bibnamefont{Schuller}}, \bibinfo{journal}{Phys. Rev. Lett.}
  \textbf{\bibinfo{volume}{84}}, \bibinfo{pages}{3466} (\bibinfo{year}{2000}).

\bibitem{hou00}
\bibinfo{author}{\bibfnamefont{C.}~\bibnamefont{Hou}},
  \bibinfo{author}{\bibfnamefont{H.}~\bibnamefont{Fujiwara}}, \bibnamefont{and}
  \bibinfo{author}{\bibfnamefont{K.}~\bibnamefont{Zhang}},
  \bibinfo{journal}{Appl. Phys. Lett} \textbf{\bibinfo{volume}{76}},
  \bibinfo{pages}{3974} (\bibinfo{year}{2000}), \bibinfo{note}{and references
  therein.}

\bibitem{yao96}
\bibinfo{author}{\bibfnamefont{Y.~D.} \bibnamefont{Yao}},
  \bibinfo{author}{\bibfnamefont{Y.~Y.} \bibnamefont{Chen}},
  \bibinfo{author}{\bibfnamefont{M.~F.} \bibnamefont{Tai}},
  \bibinfo{author}{\bibfnamefont{D.~H.} \bibnamefont{Wang}}, \bibnamefont{and}
  \bibinfo{author}{\bibfnamefont{H.~M.} \bibnamefont{Lin}},
  \bibinfo{journal}{Mater. Sci. Eng A} \textbf{\bibinfo{volume}{217-218}},
  \bibinfo{pages}{281} (\bibinfo{year}{1996}).

\bibitem{papa90}
\bibinfo{author}{\bibfnamefont{V.}~\bibnamefont{Papaefthymiou}},
  \bibinfo{author}{\bibfnamefont{A.}~\bibnamefont{Kostikas}},
  \bibinfo{author}{\bibfnamefont{A.}~\bibnamefont{Simopulos}},
  \bibinfo{author}{\bibfnamefont{D.}~\bibnamefont{Niarchos}},
  \bibinfo{author}{\bibfnamefont{S.}~\bibnamefont{Gangopadhyay}},
  \bibinfo{author}{\bibfnamefont{G.~C.} \bibnamefont{Hadjipanayis}},
  \bibinfo{author}{\bibfnamefont{C.~M.} \bibnamefont{Sorensen}},
  \bibnamefont{and} \bibinfo{author}{\bibfnamefont{K.~J.}
  \bibnamefont{Klabunde}}, \bibinfo{journal}{J. Appl. Phys.}
  \textbf{\bibinfo{volume}{67}}, \bibinfo{pages}{4487} (\bibinfo{year}{1990}).

\bibitem{lin95}
\bibinfo{author}{\bibfnamefont{H.~M.} \bibnamefont{Lin}},
  \bibinfo{author}{\bibfnamefont{C.~M.} \bibnamefont{Hsu}},
  \bibinfo{author}{\bibfnamefont{Y.~D.} \bibnamefont{Yao}},
  \bibinfo{author}{\bibfnamefont{Y.~Y.} \bibnamefont{Chen}},
  \bibinfo{author}{\bibfnamefont{T.~T.} \bibnamefont{Kuan}},
  \bibinfo{author}{\bibfnamefont{F.~A.} \bibnamefont{Kuan}},
  \bibinfo{author}{\bibfnamefont{F.~A.} \bibnamefont{Yang}}, \bibnamefont{and}
  \bibinfo{author}{\bibfnamefont{C.~Y.} \bibnamefont{Tung}},
  \bibinfo{journal}{NanoStruct. Mater.} \textbf{\bibinfo{volume}{6}},
  \bibinfo{pages}{977} (\bibinfo{year}{1995}).

\bibitem{hsu94}
\bibinfo{author}{\bibfnamefont{C.~M.} \bibnamefont{Hsu}},
  \bibinfo{author}{\bibfnamefont{H.~M.} \bibnamefont{Lin}}, \bibnamefont{and}
  \bibinfo{author}{\bibfnamefont{K.~R.} \bibnamefont{Tsai}},
  \bibinfo{journal}{J. Appl. Phys.} \textbf{\bibinfo{volume}{76}},
  \bibinfo{pages}{4793} (\bibinfo{year}{1994}).

\bibitem{sort99}
\bibinfo{author}{\bibfnamefont{J.}~\bibnamefont{Sort}},
  \bibinfo{author}{\bibfnamefont{J.}~\bibnamefont{Nogu{\'e}s}},
  \bibinfo{author}{\bibfnamefont{X.}~\bibnamefont{Amils}},
  \bibinfo{author}{\bibfnamefont{S.}~\bibnamefont{Suri{\~n}ach}},
  \bibinfo{author}{\bibfnamefont{J.~S.} \bibnamefont{Mu{\~n}oz}},
  \bibnamefont{and} \bibinfo{author}{\bibfnamefont{M.~D.}
  \bibnamefont{Bar{\'o}}}, \bibinfo{journal}{J. Appl. Phys.}
  \textbf{\bibinfo{volume}{67}}, \bibinfo{pages}{3177} (\bibinfo{year}{1999}).

\bibitem{nogu96b}
\bibinfo{author}{\bibfnamefont{J.}~\bibnamefont{Nogu{\'e}s}},
  \bibinfo{author}{\bibfnamefont{D.}~\bibnamefont{Lederman}},
  \bibinfo{author}{\bibfnamefont{T.~J.} \bibnamefont{Moran}},
  \bibinfo{author}{\bibfnamefont{I.~K.} \bibnamefont{Schuller}},
  \bibnamefont{and} \bibinfo{author}{\bibfnamefont{K.~V.} \bibnamefont{Rao}},
  \bibinfo{journal}{Appl. Phys. Lett.} \textbf{\bibinfo{volume}{68}},
  \bibinfo{pages}{3186} (\bibinfo{year}{1996}).

\bibitem{jung94}
\bibinfo{author}{\bibfnamefont{R.}~\bibnamefont{Jungblut}},
  \bibinfo{author}{\bibfnamefont{R.}~\bibnamefont{Coehoorn}},
  \bibinfo{author}{\bibfnamefont{M.~T.} \bibnamefont{Johnson}},
  \bibinfo{author}{\bibfnamefont{J.}~\bibnamefont{van~de Stegge}},
  \bibnamefont{and} \bibinfo{author}{\bibfnamefont{A.}~\bibnamefont{Reinders}},
  \bibinfo{journal}{J. Appl. Phys.} \textbf{\bibinfo{volume}{75}},
  \bibinfo{pages}{6659} (\bibinfo{year}{1994}).

\bibitem{dien91a}
\bibinfo{author}{\bibfnamefont{B.}~\bibnamefont{Dieny}},
  \bibinfo{author}{\bibfnamefont{V.~S.} \bibnamefont{Speriosu}},
  \bibinfo{author}{\bibfnamefont{S.}~\bibnamefont{Metin}},
  \bibinfo{author}{\bibfnamefont{S.~S.~P.} \bibnamefont{Parkin}},
  \bibinfo{author}{\bibfnamefont{B.~A.} \bibnamefont{Gurney}},
  \bibinfo{author}{\bibfnamefont{P.}~\bibnamefont{Baumgart}}, \bibnamefont{and}
  \bibinfo{author}{\bibfnamefont{D.~R.} \bibnamefont{Wilhoit}},
  \bibinfo{journal}{J. Appl. Phys.} \textbf{\bibinfo{volume}{69}},
  \bibinfo{pages}{4774} (\bibinfo{year}{1991}).

\bibitem{dien91b}
\bibinfo{author}{\bibfnamefont{B.}~\bibnamefont{Dieny}},
  \bibinfo{author}{\bibfnamefont{V.~S.} \bibnamefont{Speriosu}},
  \bibinfo{author}{\bibfnamefont{S.~S.~P.} \bibnamefont{Parkin}},
  \bibinfo{author}{\bibfnamefont{B.~A.} \bibnamefont{Gurney}},
  \bibinfo{author}{\bibfnamefont{D.~R.} \bibnamefont{Wilhoit}},
  \bibnamefont{and} \bibinfo{author}{\bibfnamefont{D.}~\bibnamefont{Mauri}},
  \bibinfo{journal}{Phys. Rev. B} \textbf{\bibinfo{volume}{43}},
  \bibinfo{pages}{1297} (\bibinfo{year}{1991}).

\bibitem{dien95}
\bibinfo{author}{\bibfnamefont{B.}~\bibnamefont{Dieny}},
  \bibinfo{author}{\bibfnamefont{A.}~\bibnamefont{Granovsky}},
  \bibinfo{author}{\bibfnamefont{A.}~\bibnamefont{Vedyaev}},
  \bibinfo{author}{\bibfnamefont{N.}~\bibnamefont{Ryzhanova}},
  \bibinfo{author}{\bibfnamefont{C.}~\bibnamefont{Comache}}, \bibnamefont{and}
  \bibinfo{author}{\bibfnamefont{L.}~\bibnamefont{Pereira}},
  \bibinfo{journal}{J. Magn. Magn. Mater.} \textbf{\bibinfo{volume}{151}},
  \bibinfo{pages}{378} (\bibinfo{year}{1995}).

\bibitem{nogu99}
\bibinfo{author}{\bibfnamefont{J.}~\bibnamefont{Nogu{\'e}s}},
  \bibinfo{author}{\bibfnamefont{T.~J.} \bibnamefont{Moran}},
  \bibinfo{author}{\bibfnamefont{D.}~\bibnamefont{Lederman}}, \bibnamefont{and}
  \bibinfo{author}{\bibfnamefont{I.~K.} \bibnamefont{Schuller}},
  \bibinfo{journal}{Phys. Rev. B} \textbf{\bibinfo{volume}{59}},
  \bibinfo{pages}{6984} (\bibinfo{year}{1999}).

\bibitem{goke01}
\bibinfo{author}{\bibfnamefont{N.~J.} \bibnamefont{G{\"o}kemeijer}},
  \bibinfo{author}{\bibfnamefont{R.~L.} \bibnamefont{Penn}},
  \bibinfo{author}{\bibfnamefont{D.~R.} \bibnamefont{Veblen}},
  \bibnamefont{and} \bibinfo{author}{\bibfnamefont{C.~L.} \bibnamefont{Chien}},
  \bibinfo{journal}{Phys. Rev. B} \textbf{\bibinfo{volume}{63}},
  \bibinfo{pages}{174422} (\bibinfo{year}{2001}).

\bibitem{mora98}
\bibinfo{author}{\bibfnamefont{T.~J.} \bibnamefont{Moran}},
  \bibinfo{author}{\bibfnamefont{J.}~\bibnamefont{Nogu{\'e}s}},
  \bibinfo{author}{\bibfnamefont{D.}~\bibnamefont{Lederman}}, \bibnamefont{and}
  \bibinfo{author}{\bibfnamefont{I.~K.} \bibnamefont{Schuller}},
  \bibinfo{journal}{Appl. Phys. Lett.} \textbf{\bibinfo{volume}{72}},
  \bibinfo{pages}{617} (\bibinfo{year}{1998}).

\bibitem{ijir98}
\bibinfo{author}{\bibfnamefont{Y.}~\bibnamefont{Ijiri}},
  \bibinfo{author}{\bibfnamefont{J.~A.} \bibnamefont{Borchers}},
  \bibinfo{author}{\bibfnamefont{R.~W.} \bibnamefont{Erwin}},
  \bibinfo{author}{\bibfnamefont{S.~H.} \bibnamefont{Lee}},
  \bibinfo{author}{\bibfnamefont{P.~J.} \bibnamefont{van~der Zaag}},
  \bibnamefont{and} \bibinfo{author}{\bibfnamefont{R.~M.} \bibnamefont{Wolf}},
  \bibinfo{journal}{Phys. Rev. Lett.} \textbf{\bibinfo{volume}{80}},
  \bibinfo{pages}{608} (\bibinfo{year}{1998}).

\bibitem{nogu00}
\bibinfo{author}{\bibfnamefont{J.}~\bibnamefont{Nogu{\'e}s}},
  \bibinfo{author}{\bibfnamefont{L.}~\bibnamefont{Morellon}},
  \bibinfo{author}{\bibfnamefont{C.}~\bibnamefont{Leighton}},
  \bibinfo{author}{\bibfnamefont{M.~R.} \bibnamefont{Ibarra}},
  \bibnamefont{and} \bibinfo{author}{\bibfnamefont{I.~K.}
  \bibnamefont{Schuller}}, \bibinfo{journal}{Phys. Rev. B}
  \textbf{\bibinfo{volume}{61}}, \bibinfo{pages}{{R6455}}
  (\bibinfo{year}{2000}).

\bibitem{leig99}
\bibinfo{author}{\bibfnamefont{C.}~\bibnamefont{Leighton}},
  \bibinfo{author}{\bibfnamefont{J.}~\bibnamefont{Nogu{\'e}s}},
  \bibinfo{author}{\bibfnamefont{H.}~\bibnamefont{Suhl}}, \bibnamefont{and}
  \bibinfo{author}{\bibfnamefont{I.~K.} \bibnamefont{Schuller}},
  \bibinfo{journal}{Phys. Rev. B} \textbf{\bibinfo{volume}{60}},
  \bibinfo{pages}{12 837} (\bibinfo{year}{1999}).

\bibitem{fitz00}
\bibinfo{author}{\bibfnamefont{M.~R.} \bibnamefont{Fitzsimmons}},
  \bibinfo{author}{\bibfnamefont{P.}~\bibnamefont{Yashar}},
  \bibinfo{author}{\bibfnamefont{C.}~\bibnamefont{Leighton}},
  \bibinfo{author}{\bibfnamefont{I.~K.} \bibnamefont{Schuller}},
  \bibinfo{author}{\bibfnamefont{J.}~\bibnamefont{Nogu{\'e}s}},
  \bibinfo{author}{\bibfnamefont{C.~F.} \bibnamefont{Majkrzak}},
  \bibnamefont{and} \bibinfo{author}{\bibfnamefont{J.~A.} \bibnamefont{Dura}},
  \bibinfo{journal}{Phys. Rev. Lett.} \textbf{\bibinfo{volume}{84}},
  \bibinfo{pages}{3986} (\bibinfo{year}{2000}).

\bibitem{lede97}
\bibinfo{author}{\bibfnamefont{D.}~\bibnamefont{Lederman}},
  \bibinfo{author}{\bibfnamefont{J.}~\bibnamefont{Nogu{\'e}s}},
  \bibnamefont{and} \bibinfo{author}{\bibfnamefont{I.~K.}
  \bibnamefont{Schuller}}, \bibinfo{journal}{Phys. Rev. B}
  \textbf{\bibinfo{volume}{56}}, \bibinfo{pages}{2332} (\bibinfo{year}{1997}).

\bibitem{char00}
\bibinfo{author}{\bibfnamefont{G.}~\bibnamefont{Charlton}},
  \bibinfo{author}{\bibfnamefont{P.~B.} \bibnamefont{Howes}},
  \bibinfo{author}{\bibfnamefont{C.~A.} \bibnamefont{Muryn}},
  \bibinfo{author}{\bibfnamefont{H.}~\bibnamefont{Raza}},
  \bibinfo{author}{\bibfnamefont{N.}~\bibnamefont{Jones}},
  \bibinfo{author}{\bibfnamefont{J.~S.~G.} \bibnamefont{Taylor}},
  \bibinfo{author}{\bibfnamefont{C.}~\bibnamefont{Norris}},
  \bibinfo{author}{\bibfnamefont{R.}~\bibnamefont{McGrath}},
  \bibinfo{author}{\bibfnamefont{D.}~\bibnamefont{Norman}},
  \bibinfo{author}{\bibfnamefont{T.~S.} \bibnamefont{Turner}},
  \bibnamefont{and} \bibinfo{author}{\bibfnamefont{G.}~\bibnamefont{Thornton}},
  \bibinfo{journal}{Phys. Rev. B} \textbf{\bibinfo{volume}{61}},
  \bibinfo{pages}{16 117} (\bibinfo{year}{2000}).

\bibitem{neel67}
\bibinfo{author}{\bibfnamefont{L.}~\bibnamefont{N\'eel}},
  \bibinfo{journal}{Ann. Phys. (Paris)} \textbf{\bibinfo{volume}{2}},
  \bibinfo{pages}{61} (\bibinfo{year}{1967}).

\bibitem{neel88}
\bibinfo{editor}{\bibfnamefont{N.}~\bibnamefont{Kurti}}, ed.,
  \emph{\bibinfo{title}{Selected Works of Louis N\'eel}}
  (\bibinfo{publisher}{Gordon and Breach}, \bibinfo{address}{New York},
  \bibinfo{year}{1988}), \bibinfo{note}{includes an english translation of the
  preceding reference.}

\bibitem{mauri87}
\bibinfo{author}{\bibfnamefont{D.}~\bibnamefont{Mauri}},
  \bibinfo{author}{\bibfnamefont{H.~C.} \bibnamefont{Siegmann}},
  \bibinfo{author}{\bibfnamefont{P.~S.} \bibnamefont{Bagus}}, \bibnamefont{and}
  \bibinfo{author}{\bibfnamefont{E.}~\bibnamefont{Kay}}, \bibinfo{journal}{J.
  Appl. Phys.} \textbf{\bibinfo{volume}{62}}, \bibinfo{pages}{3047}
  (\bibinfo{year}{1987}).

\bibitem{malo87}
\bibinfo{author}{\bibfnamefont{A.~P.} \bibnamefont{Malozemoff}},
  \bibinfo{journal}{Phys. Rev. B} \textbf{\bibinfo{volume}{35}},
  \bibinfo{pages}{3679} (\bibinfo{year}{1987}).

\bibitem{malo88a}
\bibinfo{author}{\bibfnamefont{A.~P.} \bibnamefont{Malozemoff}},
  \bibinfo{journal}{Phys. Rev. B} \textbf{\bibinfo{volume}{37}},
  \bibinfo{pages}{7673} (\bibinfo{year}{1988}).

\bibitem{malo88b}
\bibinfo{author}{\bibfnamefont{A.~P.} \bibnamefont{Malozemoff}},
  \bibinfo{journal}{J. Appl. Phys.} \textbf{\bibinfo{volume}{63}},
  \bibinfo{pages}{3874} (\bibinfo{year}{1988}).

\bibitem{koon97}
\bibinfo{author}{\bibfnamefont{N.~C.} \bibnamefont{Koon}},
  \bibinfo{journal}{Phys. Rev. Lett.} \textbf{\bibinfo{volume}{78}},
  \bibinfo{pages}{4865} (\bibinfo{year}{1997}).

\bibitem{schu98}
\bibinfo{author}{\bibfnamefont{T.~C.} \bibnamefont{Schulthess}}
  \bibnamefont{and} \bibinfo{author}{\bibfnamefont{W.~H.}
  \bibnamefont{Butler}}, \bibinfo{journal}{Phys. Rev. Lett.}
  \textbf{\bibinfo{volume}{81}}, \bibinfo{pages}{4516} (\bibinfo{year}{1998}).

\bibitem{schu99}
\bibinfo{author}{\bibfnamefont{T.~C.} \bibnamefont{Schulthess}}
  \bibnamefont{and} \bibinfo{author}{\bibfnamefont{W.~H.}
  \bibnamefont{Butler}}, \bibinfo{journal}{J. Appl. Phys.}
  \textbf{\bibinfo{volume}{85}}, \bibinfo{pages}{5510} (\bibinfo{year}{1999}).

\bibitem{stil99}
\bibinfo{author}{\bibfnamefont{M.~D.} \bibnamefont{Stiles}} \bibnamefont{and}
  \bibinfo{author}{\bibfnamefont{R.~D.} \bibnamefont{McMichael}},
  \bibinfo{journal}{Phys. Rev. B} \textbf{\bibinfo{volume}{59}},
  \bibinfo{pages}{3722} (\bibinfo{year}{1999}).

\bibitem{kmpr1}
\bibinfo{author}{\bibfnamefont{M.}~\bibnamefont{Kiwi}},
  \bibinfo{author}{\bibfnamefont{J.}~\bibnamefont{Mej\'{\i}a-L{\'o}pez}},
  \bibinfo{author}{\bibfnamefont{R.~D.} \bibnamefont{Portugal}},
  \bibnamefont{and}
  \bibinfo{author}{\bibfnamefont{R.}~\bibnamefont{Ram\'{\i}rez}},
  \bibinfo{journal}{Europhys. Lett} \textbf{\bibinfo{volume}{48}},
  \bibinfo{pages}{573} (\bibinfo{year}{1999}).

\bibitem{kmpr2}
\bibinfo{author}{\bibfnamefont{M.}~\bibnamefont{Kiwi}},
  \bibinfo{author}{\bibfnamefont{J.}~\bibnamefont{Mej\'{\i}a-L{\'o}pez}},
  \bibinfo{author}{\bibfnamefont{R.~D.} \bibnamefont{Portugal}},
  \bibnamefont{and}
  \bibinfo{author}{\bibfnamefont{R.}~\bibnamefont{Ram\'{\i}rez}},
  \bibinfo{journal}{Appl. Phys. Lett.} \textbf{\bibinfo{volume}{75}},
  \bibinfo{pages}{3995} (\bibinfo{year}{1999}).

\bibitem{kmpr3}
\bibinfo{author}{\bibfnamefont{M.}~\bibnamefont{Kiwi}},
  \bibinfo{author}{\bibfnamefont{J.}~\bibnamefont{Mej\'{\i}a-L{\'o}pez}},
  \bibinfo{author}{\bibfnamefont{R.~D.} \bibnamefont{Portugal}},
  \bibnamefont{and}
  \bibinfo{author}{\bibfnamefont{R.}~\bibnamefont{Ram\'{\i}rez}},
  \bibinfo{journal}{Solid State Comm.} \textbf{\bibinfo{volume}{116}},
  \bibinfo{pages}{315} (\bibinfo{year}{2000}).

\bibitem{taka97}
\bibinfo{author}{\bibfnamefont{K.}~\bibnamefont{Takano}},
  \bibinfo{author}{\bibfnamefont{R.~H.} \bibnamefont{Kodama}},
  \bibinfo{author}{\bibfnamefont{A.~E.} \bibnamefont{Berkowitz}},
  \bibinfo{author}{\bibfnamefont{W.}~\bibnamefont{Cao}}, \bibnamefont{and}
  \bibinfo{author}{\bibfnamefont{G.}~\bibnamefont{Thomas}},
  \bibinfo{journal}{Phys. Rev. Lett.} \textbf{\bibinfo{volume}{79}},
  \bibinfo{pages}{1130} (\bibinfo{year}{1997}).

\bibitem{zhan99}
\bibinfo{author}{\bibfnamefont{S.}~\bibnamefont{Zhang}},
  \bibinfo{author}{\bibfnamefont{D.}~\bibnamefont{Dimitrov}},
  \bibinfo{author}{\bibfnamefont{G.~C.} \bibnamefont{Hadjipanayis}},
  \bibinfo{author}{\bibfnamefont{J.~W.} \bibnamefont{Cai}}, \bibnamefont{and}
  \bibinfo{author}{\bibfnamefont{C.~L.} \bibnamefont{Chien}},
  \bibinfo{journal}{J. Magn. Magn. Mat.} \textbf{\bibinfo{volume}{198-199}},
  \bibinfo{pages}{468} (\bibinfo{year}{1999}).

\bibitem{dimi98}
\bibinfo{author}{\bibfnamefont{D.~V.} \bibnamefont{Dimitrov}},
  \bibinfo{author}{\bibfnamefont{S.}~\bibnamefont{Zhang}},
  \bibinfo{author}{\bibfnamefont{J.~Q.} \bibnamefont{Xiao}},
  \bibinfo{author}{\bibfnamefont{G.~C.} \bibnamefont{Hadjipanayis}},
  \bibnamefont{and} \bibinfo{author}{\bibfnamefont{C.}~\bibnamefont{Prados}},
  \bibinfo{journal}{Phys. Rev B} \textbf{\bibinfo{volume}{58}},
  \bibinfo{pages}{12090} (\bibinfo{year}{1998}).

\bibitem{meik62}
\bibinfo{author}{\bibfnamefont{W.~P.} \bibnamefont{Meiklejohn}},
  \bibinfo{journal}{J. Appl. Phys.} \textbf{\bibinfo{volume}{33}},
  \bibinfo{pages}{1328} (\bibinfo{year}{1962}).

\bibitem{kurz01}
\bibinfo{author}{\bibfnamefont{P.}~\bibnamefont{Kurz}},
  \bibinfo{author}{\bibfnamefont{G.}~\bibnamefont{Bihlmayer}},
  \bibinfo{author}{\bibfnamefont{K.}~\bibnamefont{Hirai}}, \bibnamefont{and}
  \bibinfo{author}{\bibfnamefont{S.}~\bibnamefont{Bl{\"u}gel}},
  \bibinfo{journal}{Phys. Rev. Lett.} \textbf{\bibinfo{volume}{86}},
  \bibinfo{pages}{1106} (\bibinfo{year}{2001}).

\bibitem{kittel_ISSP}
\bibinfo{author}{\bibfnamefont{C.}~\bibnamefont{Kittel}},
  \emph{\bibinfo{title}{Introduction to Solid State Physics}}
  (\bibinfo{publisher}{John Wiley \& Sons}, \bibinfo{address}{New York},
  \bibinfo{year}{1986}), {S}ixth ed.

\bibitem{nogu00+}
\bibinfo{author}{\bibfnamefont{J.}~\bibnamefont{Nogu{\'e}s}},
  \bibinfo{author}{\bibfnamefont{C.}~\bibnamefont{Leighton}}, \bibnamefont{and}
  \bibinfo{author}{\bibfnamefont{I.~K.} \bibnamefont{Schuller}},
  \bibinfo{journal}{Phys. Rev. B} \textbf{\bibinfo{volume}{61}},
  \bibinfo{pages}{{1315}} (\bibinfo{year}{2000}).

\bibitem{mill96}
\bibinfo{author}{\bibfnamefont{B.~H.} \bibnamefont{Miller}} \bibnamefont{and}
  \bibinfo{author}{\bibfnamefont{E.~D.} \bibnamefont{Dahlberg}},
  \bibinfo{journal}{Appl. Phys. Lett.} \textbf{\bibinfo{volume}{69}},
  \bibinfo{pages}{3932} (\bibinfo{year}{1996}).

\bibitem{stro97}
\bibinfo{author}{\bibfnamefont{V.}~\bibnamefont{Str{\"o}m}},
  \bibinfo{author}{\bibfnamefont{B.~J.} \bibnamefont{J{\"o}nsson}},
  \bibinfo{author}{\bibfnamefont{K.~V.} \bibnamefont{Rao}}, \bibnamefont{and}
  \bibinfo{author}{\bibfnamefont{E.~D.} \bibnamefont{Dahlberg}},
  \bibinfo{journal}{J. Appl. Phys.} \textbf{\bibinfo{volume}{81}},
  \bibinfo{pages}{5003} (\bibinfo{year}{1997}).

\bibitem{nogu99b}
\bibinfo{author}{\bibfnamefont{J.}~\bibnamefont{Nogu{\'e}s}} \bibnamefont{and}
  \bibinfo{author}{\bibfnamefont{I.~K.} \bibnamefont{Schuller}},
  \bibinfo{journal}{J. Magn. Magn. Mat.} \textbf{\bibinfo{volume}{192}},
  \bibinfo{pages}{203} (\bibinfo{year}{1999}).

\bibitem{full98}
\bibinfo{author}{\bibfnamefont{E.~E.} \bibnamefont{Fullerton}},
  \bibinfo{author}{\bibfnamefont{J.~S.} \bibnamefont{Jiang}},
  \bibinfo{author}{\bibfnamefont{M.}~\bibnamefont{Grimsditch}},
  \bibinfo{author}{\bibfnamefont{C.~H.} \bibnamefont{Sowers}},
  \bibnamefont{and} \bibinfo{author}{\bibfnamefont{S.~D.} \bibnamefont{Bader}},
  \bibinfo{journal}{Phys. Rev. B} \textbf{\bibinfo{volume}{58}},
  \bibinfo{pages}{12193} (\bibinfo{year}{1998}).

\bibitem{labo69}
\bibinfo{author}{\bibfnamefont{A.~E.} \bibnamefont{LaBonte}} \bibnamefont{and}
  \bibinfo{author}{\bibfnamefont{J.~S.} \bibnamefont{Kouvel}},
  \bibinfo{journal}{J. Appl. Phys.} \textbf{\bibinfo{volume}{40}},
  \bibinfo{pages}{2450} (\bibinfo{year}{1969}).

\bibitem{brow78}
\bibinfo{author}{\bibfnamefont{J.}~\bibnamefont{William F.~Brown}},
  \emph{\bibinfo{title}{Micromagnetics}} (\bibinfo{publisher}{Robert E. Krieger
  Publ. Co.}, \bibinfo{address}{Huntington, New York}, \bibinfo{year}{1978}).

\bibitem{ahar96}
\bibinfo{author}{\bibfnamefont{A.}~\bibnamefont{Aharoni}},
  \emph{\bibinfo{title}{Introduction to the Theory of Ferromagnetism}}
  (\bibinfo{publisher}{Oxford Science Publications}, \bibinfo{address}{Oxford,
  England}, \bibinfo{year}{1996}), \bibinfo{note}{and references therein.}

\bibitem{mora95}
\bibinfo{author}{\bibfnamefont{T.~J.} \bibnamefont{Moran}},
  \bibinfo{author}{\bibfnamefont{J.~M.} \bibnamefont{Gallego}},
  \bibnamefont{and} \bibinfo{author}{\bibfnamefont{I.~K.}
  \bibnamefont{Schuller}}, \bibinfo{journal}{J. Appl. Phys.}
  \textbf{\bibinfo{volume}{78}}, \bibinfo{pages}{1887} (\bibinfo{year}{1995}).

\bibitem{slon91}
\bibinfo{author}{\bibfnamefont{J.~C.} \bibnamefont{Slonczewski}},
  \bibinfo{journal}{Phys. Rev. Lett.} \textbf{\bibinfo{volume}{67}},
  \bibinfo{pages}{3172} (\bibinfo{year}{1991}).

\bibitem{carr94}
\bibinfo{author}{\bibfnamefont{A.~S.} \bibnamefont{Carri{\c c}o}},
  \bibinfo{author}{\bibfnamefont{R.~E.} \bibnamefont{Camley}},
  \bibnamefont{and} \bibinfo{author}{\bibfnamefont{R.~L.}
  \bibnamefont{Stamps}}, \bibinfo{journal}{Phys. Rev. B}
  \textbf{\bibinfo{volume}{50}}, \bibinfo{pages}{13453} (\bibinfo{year}{1994}).

\bibitem{caml87}
\bibinfo{author}{\bibfnamefont{R.~E.} \bibnamefont{Camley}},
  \bibinfo{journal}{Phys. Rev. B} \textbf{\bibinfo{volume}{35}},
  \bibinfo{pages}{3608} (\bibinfo{year}{1987}).

\bibitem{caml88}
\bibinfo{author}{\bibfnamefont{R.~E.} \bibnamefont{Camley}} \bibnamefont{and}
  \bibinfo{author}{\bibfnamefont{D.~R.} \bibnamefont{Tilley}},
  \bibinfo{journal}{Phys. Rev. B} \textbf{\bibinfo{volume}{37}},
  \bibinfo{pages}{3413} (\bibinfo{year}{1988}).

\bibitem{kirk83}
\bibinfo{author}{\bibfnamefont{S.}~\bibnamefont{Kirkpatrick}},
  \bibinfo{author}{\bibfnamefont{C.~D.} \bibnamefont{Gelatt}},
  \bibnamefont{and} \bibinfo{author}{\bibfnamefont{M.~P.} \bibnamefont{Vechi}},
  \bibinfo{journal}{Science} \textbf{\bibinfo{volume}{220}},
  \bibinfo{pages}{671} (\bibinfo{year}{1983}).

\bibitem{full98b}
\bibinfo{author}{\bibfnamefont{E.~E.} \bibnamefont{Fullerton}},
  \bibinfo{author}{\bibfnamefont{J.~S.} \bibnamefont{Jiang}},
  \bibinfo{author}{\bibfnamefont{C.~H.} \bibnamefont{Sowers}},
  \bibinfo{author}{\bibfnamefont{J.~E.} \bibnamefont{Pearson}},
  \bibnamefont{and} \bibinfo{author}{\bibfnamefont{S.~D.} \bibnamefont{Bader}},
  \bibinfo{journal}{Appl. Phys. Lett.} \textbf{\bibinfo{volume}{72}},
  \bibinfo{pages}{380} (\bibinfo{year}{1998}).

\bibitem{nolt00}
\bibinfo{author}{\bibfnamefont{F.}~\bibnamefont{Nolting}},
  \bibinfo{author}{\bibfnamefont{A.}~\bibnamefont{Scholl}},
  \bibinfo{author}{\bibfnamefont{J.}~\bibnamefont{St{\"o}hr}},
  \bibinfo{author}{\bibfnamefont{J.~W.} \bibnamefont{Seo}},
  \bibinfo{author}{\bibfnamefont{J.}~\bibnamefont{Fompeyrine}},
  \bibinfo{author}{\bibfnamefont{H.}~\bibnamefont{Siegwart}},
  \bibinfo{author}{\bibfnamefont{J.-P.} \bibnamefont{Locquet}},
  \bibinfo{author}{\bibfnamefont{S.}~\bibnamefont{Anders}},
  \bibinfo{author}{\bibfnamefont{J.}~\bibnamefont{L{\"u}ning}},
  \bibinfo{author}{\bibfnamefont{E.~E.} \bibnamefont{Fullerton}},
  \bibinfo{author}{\bibfnamefont{M.~F.} \bibnamefont{Toney}},
  \bibinfo{author}{\bibfnamefont{M.~R.} \bibnamefont{Scheinfein}},
  \emph{et~al.}, \bibinfo{journal}{Nature} \textbf{\bibinfo{volume}{405}},
  \bibinfo{pages}{{767}} (\bibinfo{year}{2000}).

\bibitem{mats00}
\bibinfo{author}{\bibfnamefont{H.}~\bibnamefont{Matsuyama}},
  \bibinfo{author}{\bibfnamefont{C.}~\bibnamefont{Hanigoya}}, \bibnamefont{and}
  \bibinfo{author}{\bibfnamefont{K.}~\bibnamefont{Koike}},
  \bibinfo{journal}{Phys. Rev. Lett.} \textbf{\bibinfo{volume}{85}},
  \bibinfo{pages}{{646}} (\bibinfo{year}{2000}).

\bibitem{hong98}
\bibinfo{author}{\bibfnamefont{T.~M.} \bibnamefont{Hong}},
  \bibinfo{journal}{Phys. Rev. B} \textbf{\bibinfo{volume}{58}},
  \bibinfo{pages}{97} (\bibinfo{year}{1998}).

\bibitem{suhl98}
\bibinfo{author}{\bibfnamefont{H.}~\bibnamefont{Suhl}} \bibnamefont{and}
  \bibinfo{author}{\bibfnamefont{I.~K.} \bibnamefont{Schuller}},
  \bibinfo{journal}{Phys. Rev. B} \textbf{\bibinfo{volume}{58}},
  \bibinfo{pages}{258} (\bibinfo{year}{1998}).

\end{thebibliography}

\end{document}